\documentclass[sigconf]{acmart}
\AtBeginDocument{%
  \providecommand\BibTeX{{%
    \normalfont B\kern-0.5em{\scshape i\kern-0.25em b}\kern-0.8em\TeX}}}

\usepackage{booktabs} 
\usepackage{multirow}
\usepackage{latexsym}
\usepackage{graphicx} 
\usepackage{algorithm}
\usepackage{algorithmic}
\usepackage{epsfig}
\usepackage{color}
\usepackage{amsmath}
\usepackage{graphicx,subfigure}
\usepackage{makecell}
\usepackage{textcomp}
\usepackage{url}
\usepackage{flushend}
\usepackage{bbm}

\acmDOI{10.475/123_4}

\acmISBN{123-4567-24-567/08/06}

\copyrightyear{2022}
\acmYear{2022}
\setcopyright{acmcopyright}
\acmConference[SIGIR '22] {Proceedings of the 45th International ACM SIGIR Conference on Research and Development in Information Retrieval}{July 11--15, 2022}{Madrid, Spain.}
\acmBooktitle{Proceedings of the 45th International ACM SIGIR Conference on Research and Development in Information Retrieval (SIGIR '22), July 11--15, 2022, Madrid, Spain}
\acmPrice{15.00}
\acmISBN{978-1-4503-8732-3/22/07}
\acmDOI{10.1145/XXXXXX.XXXXXX}

\begin{document}
\title{ProFairRec: Provider Fairness-aware News Recommendation}

\author{Tao Qi}
\affiliation{%
  \institution{Department of Electronic Engineering, Tsinghua University}
  \country{}
}
\email{taoqi.qt@gmail.com}

\author{Fangzhao Wu}
\authornote{The corresponding author.}
\affiliation{%
  \institution{Microsoft Research Asia}
  \country{}
}
\email{wufangzhao@gmail.com}

\author{Chuhan Wu}
\affiliation{%
  \institution{Department of Electronic Engineering, Tsinghua University}
  \country{}
}
\email{wuchuhan15@gmail.com}

\author{Peijie Sun}
\affiliation{%
  \institution{School of Computer Science and Information Engineering, Hefei University of Technology}
  \country{}
}
\email{sun.hfut@gmail.com}

\author{Le Wu}
\affiliation{%
  \institution{School of Computer Science and Information Engineering, Hefei University of Technology}
  \country{}
}
\email{lewu.ustc@gmail.com}

\author{Xiting Wang}
\affiliation{%
  \institution{Microsoft Research Asia}
  \country{}
}
\email{xitwan@microsoft.com}

\author{Yongfeng Huang}
\affiliation{%
  \institution{Department of Electronic Engineering, Tsinghua University}
  \country{}
}
\email{yfhuang@tsinghua.edu.cn}

\author{Xing Xie}
\affiliation{%
  \institution{Microsoft Research Asia}
  \country{}
}
\email{xingx@microsoft.com}

\begin{abstract}

News recommendation aims to help online news platform users find their preferred news articles.
Existing news recommendation methods usually learn models from historical user behaviors on news.
However, these behaviors are usually biased on news providers.
Models trained on biased user data may capture and even amplify the biases on news providers, and are unfair for some minority news providers.
In this paper, we propose a provider fairness-aware news recommendation framework (named \textit{ProFairRec}), which can learn news recommendation models fair for different news providers from biased user data.
The core idea of \textit{ProFairRec} is to learn provider-fair news representations and provider-fair user representations to achieve provider fairness.
To learn provider-fair representations from biased data, we employ provider-biased representations to inherit provider bias from data. 
Provider-fair and -biased news representations are learned from news content and provider IDs respectively, which are further aggregated to build fair and biased user representations based on user click history.
All of these representations are used in model training while only fair representations are used for user-news matching to achieve fair news recommendation.
Besides, we propose an adversarial learning task on news provider discrimination to prevent provider-fair news representation from encoding provider bias.
We also propose an orthogonal regularization on provider-fair and -biased representations to better reduce provider bias in provider-fair representations.
Moreover, \textit{ProFairRec} is a general framework and can be applied to different news recommendation methods.
Extensive experiments on a public dataset verify that our \textit{ProFairRec} approach can effectively improve the provider fairness of many existing methods and meanwhile maintain their recommendation accuracy.

\end{abstract}



\maketitle

\section{INTRODUCTION}
In recent years, online news platforms such as Apple News have become more and more popular for people to read news~\cite{wu2021newsbert,yi2021efficient,wutanr}.
Since massive news articles are collected by online news platforms from different news providers everyday~\cite{wuuser,liu2010personalized}, it is very difficult for users to manually find their interested news~\cite{wu2019neurald,danzhu2019}.
Thus, news recommendation, which aims to recommend news according to users' personal interest~\cite{qi2020privacy,zheng2018drn,qi2021kim}, has become very critical for online news platforms to improve user experience~\cite{bansal2015content,kompan2010content,khattar2018weave,qi2021uni}.

Most of the existing news recommendation methods first learn news and user representations from news content and user reading history respectively, and then match them for personalized news recommendation~\cite{okura2017embedding,zhang2021unbert}.
Besides, these methods usually rely on historical users' behavior data on news to train recommendation models~\cite{wang2020fine,wu2020ptum}.
For example, \citet{an2019neural} employed an attentive convolutional network to learn news representations from news titles and categories.
They proposed to capture short-term user interest from user's clicked news via a GRU network and long-term user interest via ID embeddings and combined them to form user interest representations.
Besides, they further model user interest in news based on the inner product of their representations and exploit user historical click and non-click data for model training.
\citet{wu2021uag} proposed to learn news representations from news titles, entities, and tags via transformers and proposed a heterogeneous graph pooling network to learn user representations from personalized user graphs.
Besides, they also performed interest matching via inner product and trained models on click data.

\begin{figure}
    \centering
    \resizebox{0.46\textwidth}{!}{
    \includegraphics{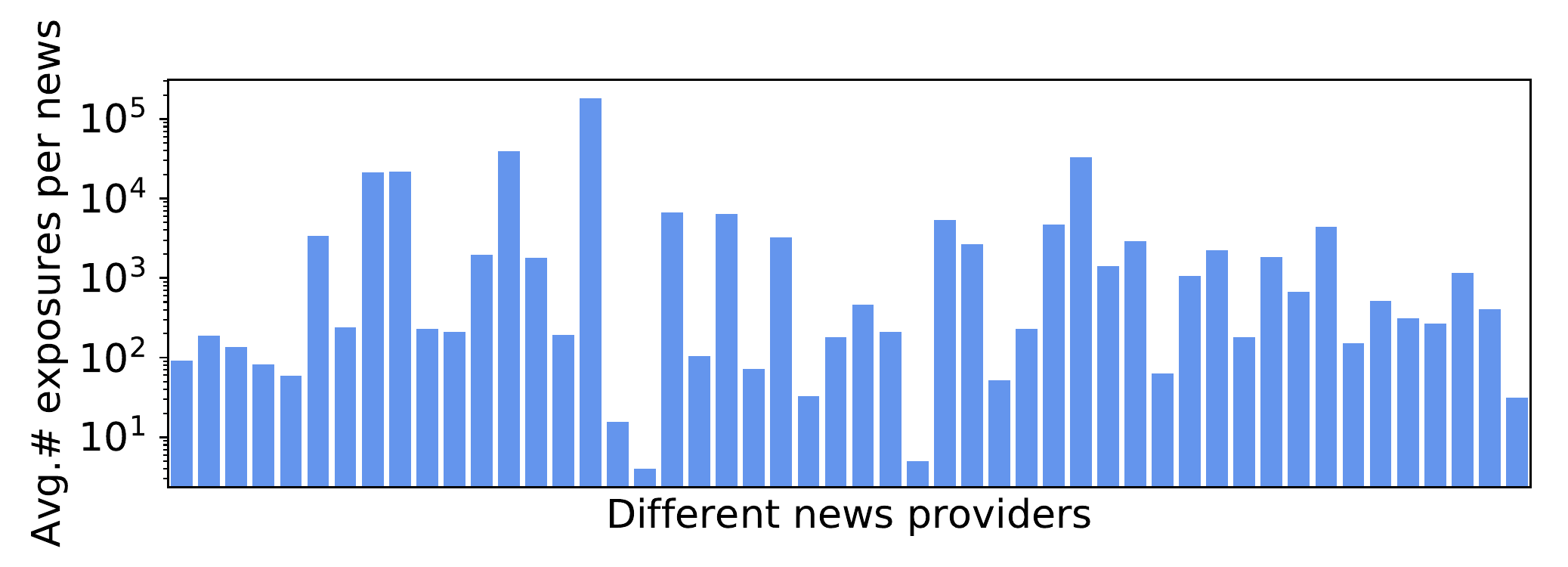}
    }
    \caption{Average exposure number per news of providers on the \textit{MIND} dataset~\cite{wu2020mind}. We randomly select 50 providers.}
    \label{fig.distribution}
\end{figure}

\begin{figure}
    \centering
    \resizebox{0.49\textwidth}{!}{
    \includegraphics{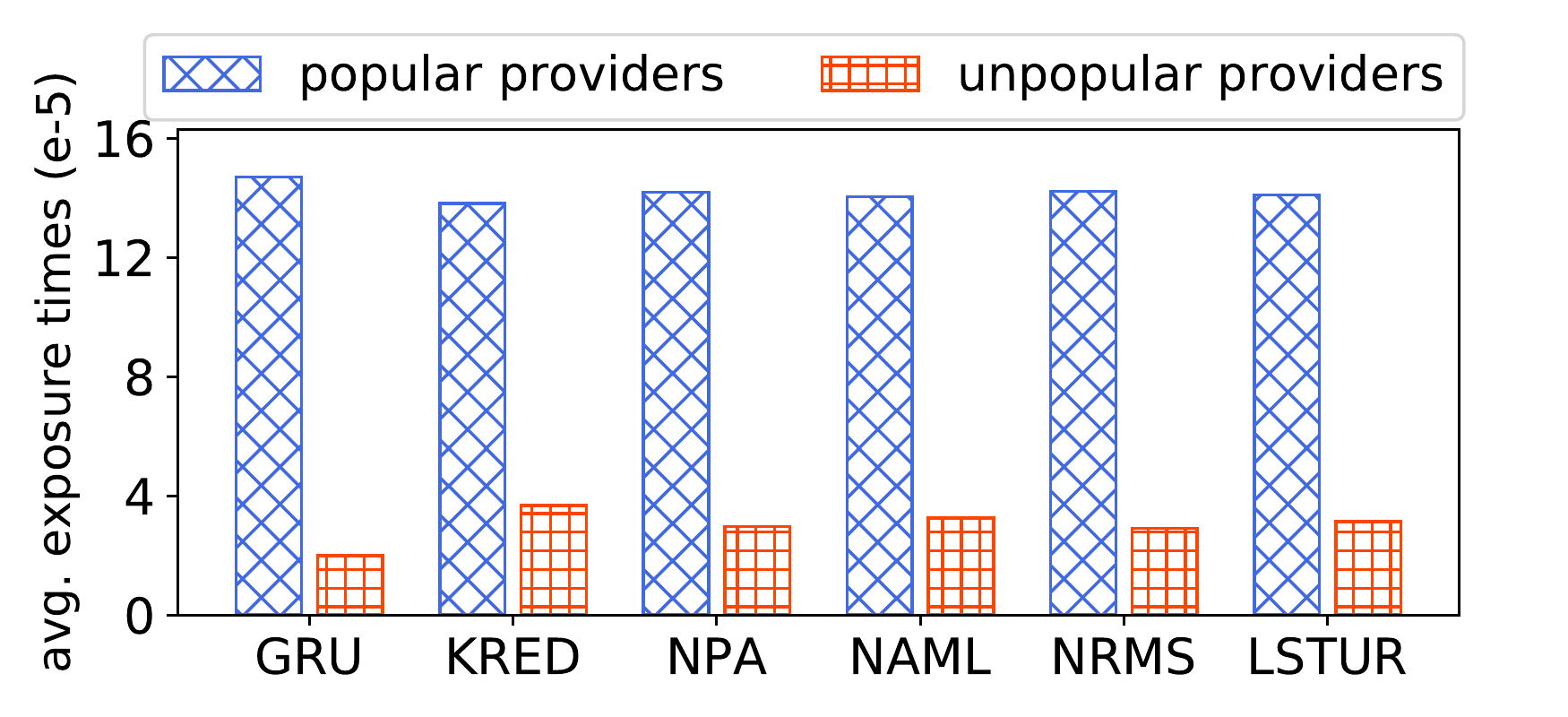}
    }
    \caption{Average exposure opportunity of popular and unpopular providers in the top recommended results of different news recommendation models on \textit{MIND}. Popular providers are defined as top 50\% providers ranked by average news click number per news article in historical data, and the remaining are regarded as unpopular providers.}
    \label{fig.unfair}
\end{figure}

In general, users' reading behaviors on different news providers are usually biased~\cite{liu2019personalized}.
Some popular news providers may take up most clicks while clicks on some minority news providers are usually rare.
Thus, news recommendation models trained on biased user data may capture and even amplify the biases on news providers, which may tend to recommend more news from popular providers and become unfair for some minority news providers~\cite{sonboli2020opportunistic,wu2021tfrom,pitoura2021fairness}.
For example, Fig.~\ref{fig.distribution} shows the average exposure number of news from different news providers in the real-world news recommendation dataset \textit{MIND}~\cite{wu2020mind}, we can see that news exposures are biased on different news providers.
Fig.~\ref{fig.unfair} shows average numbers of news from popular and unpopular providers recommended by several SOTA news recommendation models trained on \textit{MIND}.
Fig.~\ref{fig.unfair} verifies that models learned from biased data can capture provider bias and become unfair for minority providers.
However, most of the existing news recommendation methods do not consider provider fairness, which may hurt the diversity of news sources and perspectives, as well as the revenue of minority providers~\cite{trielli2019search}.

In this paper, we propose a provider fairness-aware news recommendation framework named \textit{ProFairRec}, which can improve provider fairness of recommendation models learned from biased data.
The core idea of \textit{ProFairRec} is to learn provider-fair news and user representations.
In order to learn provider-fair representations from biased data, we utilize provider-biased representations to inherit bias from data.
Specifically, for news modeling, we learn provider-fair news representations from news content and learn provider-biased news representations from provider IDs.
For user modeling, we learn provider-fair and -biased user representations by aggregating provider-fair and -biased representations of users' clicked news, respectively.
We aggregate provider-fair and -biased representations of news and users for training the recommendation model to achieve decomposition of bias-aware and bias-independent information and only match provider-fair news and user representations to achieve fair news recommendation.
In addition, we propose an adversarial learning task on news provider discrimination and apply it to provider-fair news representations to prevent them from encoding provider bias.
We also design an orthogonal regularization that enforces provider-fair and -biased representations to be orthogonal for better reducing bias in provider-fair representations.
Moreover, \textit{ProFairRec} is a general framework and can be applied to different news recommendation models.
We conduct extensive experiments on a public news recommendation dataset.
Results show that \textit{ProFairRec} can effectively improve the provider fairness of many news recommendation methods and meanwhile maintain their recommendation accuracy.

The contribution of this paper is three-fold:
\begin{itemize}

    \item To our best knowledge, we are the first to study the provider fairness problem in news recommendation.
    
    \item We propose a unified framework to learn provider-fair news representations and user representations from biased data.
    
    \item Extensive experiments verify that \textit{ProFairRec} can effectively improve provider fairness of many news recommendation methods without significant accuracy drop. 

\end{itemize}

\section{RELATED WORK}

\subsection{News Recommendation}


News recommendation technique is important for users to improve their experience on online news platforms~\cite{das2007google,lin2014personalized}.
Mainstream news recommendation methods usually first model news from its content and model user interest from user's clicked news, and further match user interest with candidate news for news recommendation~\cite{wu2021empowering,wang2011collaborative,wutanr,lian2018towards,wu2019neurald}.
Besides, these methods usually train recommendation models on historical user behavior data.
For example, \citet{wu2019neuralc} proposed to apply self-attention networks to learn news representations from news texts and learn user representations from user behavior contexts.
They further measured user interest in candidate news based on the inner product of their representations.
\citet{qi2021hierec} proposed to learn news representations from news titles and entities via transformer networks.
They further proposed to learn hierarchical user representations to model diverse and multi-grained user interest for matching candidate news.
Moreover, both of these two methods trained news recommendation models on click and non-click data.
In general, historical click data on different news providers are usually biased.
Popular providers may accumulate most clicks while click data on some minority news providers are usually rare.
However, existing news recommendation models trained on historical data may inherit provider bias and become unfair for some minority news providers.
Different from these methods, we propose a decomposed framework with adversarial learning, which can effectively reduce provider bias in news recommendation models learned from biased data.

\subsection{Fair Recommender System}

In recent years, fair recommender systems have attracted more and more attentions~\cite{ge2021towards,morik2020controlling,wu2021tfrom,li2021user}.
Existing works usually consider recommendation fairness from two perspectives, i.e., user and provider~\cite{burke2018balanced}.
Many existing fair recommender systems were proposed to improve user-side fairness~\cite{wu2021fairness,li2021towards}.
Some of these methods were designed to provide unbiased recommendations for users with different sensitive attributes (e.g., genders and ages)~\cite{wu2021fairness,wu2021learning} while some of them were proposed to provide recommendation services with equal qualities for users in different groups (e.g., user groups partitioned by gender)~\cite{li2021user,geyik2019fairness,yao2017beyond}.
Different from these user-side fair recommendation methods, our \textit{ProFairRec} framework focuses on provider-side recommendation fairness.
There are also some works that have been proposed to improve provider-side recommendation fairness~\cite{burke2018balanced,sonboli2020opportunistic}.
These methods are usually designed to provide equal recommendation opportunities for different news providers~\cite{patro2020fairrec,morik2020controlling,wu2021tfrom}.
For example, \citet{liu2019personalized} proposed to re-rank candidate items based on their relevance with user interest and proportion of protected providers (e.g., minority providers) in recommendation lists.
\citet{sonboli2020opportunistic} proposed to prompt provider fairness by enhancing recommendation diversity.
They proposed to re-rank candidate items based on their matching with user interest as well as intra-list item similarities.
\citet{wu2021tfrom} proposed to use the difference between the proportion of protected providers in the recommendation list and item set as a constraint to re-rank candidate items.
In short, these methods are usually based on manually designed re-ranking rules to re-balance the recommendation opportunities of different news providers to improve provider fairness.
However, the manually designed re-ranking rules may be sub-optimal for achieving an effective trade-off between recommendation fairness and accuracy.
Different from these methods, we propose a unified provider fairness-aware news recommendation framework to reduce provider bias in recommendation models with adversarial learning, which can jointly optimize recommendation fairness and accuracy during model training.
\section{Methodology}

We will first give a formal definition on news recommendation and provider fairness in it.
Then, we will introduce our provider fairness-aware news recommendation method (named \textit{ProFairRec}).


\subsection{Problem Formulation}

In news recommendation, we assume that there are $H$ news providers in total: $\mathcal{P}=\{P_i|i=1,...,H\}$, where $P_i$ is the ID of the $i$-th provider and $\mathcal{P}$ is the set of news providers.
The set of news articles belonging to the provider $p \in \mathcal{P}$ is denoted as $\mathcal{D}_p$ and the set of all news articles is denoted as $\mathcal{D}$.
For each news article $d \in \mathcal{D}$, we assume that the article $d$ is composed of the textual content $c$ and the ID of its provider $p$, i.e., $d=\{c,p\}$.
Besides, we assume that the textual content $c$ is composed of $T$ words: $c=[t_1,...,t_T]$, where $t_i$ is the $i$-th word.
For each target user $u \in \mathcal{U}$, we assume that the target user has previously clicked $m$ news articles $[d^u_1,...,d^u_i,...,d^u_m]$, where $d^u_i$ is the $i$-th news clicked by the target user and $\mathcal{U}$ is the set of users.
Given a list of candidate news $[d^n_1,...,d^n_i,...,d^n_N]$ and a target user $u$, the recommendation model needs to predict the relevance scores $[\hat{r}^c_1,...,\hat{r}^c_i,...,\hat{r}^c_N]$ for matching user interest and candidate news, where $\hat{r}^c_i$ is the predicted relevance score of the $i$-th candidate news and user interest.
These candidate news are further ranked based on their relevance scores and the news recommender system will only display part of candidate news with the highest relevance scores.
The goal of provider fairness-aware news recommendation is to recommend news according to user interest and meanwhile provide fair exposure opportunities for different news providers.

Following existing works~\cite{burke2018balanced,sonboli2020opportunistic}, we employ group-level fairness concept to define provider fairness to reduce the influence of the quality differences of news from different providers.
News providers are first partitioned into multiple groups $\{\mathcal{P}_i|i=1,2,...,g\}$ based on concerned provider features (e.g., provider popularity~\cite{sonboli2020opportunistic} or provider revenue~\cite{burke2018balanced}), where $\mathcal{P}_i$ is the $i$-th group and $g$ is the number of provider groups.
The group-level provider fairness requires that the recommendation model should provide equal recommendation opportunities for news of providers in different groups:
\begin{equation}
\label{eq.profair}
    \frac{\sum_{u\in\mathcal{U}} \sum_{p\in\mathcal{P}_i} \sum_{d_i\in \mathcal{D}_p} \mathbbm{1}(d_i,u) }{ |\mathcal{U}| \sum_{p\in\mathcal{P}_i} |\mathcal{D}_p|} =  \frac{\sum_{u\in\mathcal{U}}  \sum_{p\in\mathcal{P}_j}  \sum_{d_j\in \mathcal{D}_p} \mathbbm{1}(d_j,u) }{|\mathcal{U}|\sum_{p\in\mathcal{P}_j} |\mathcal{D}_p|}, 
\end{equation}
where $\mathbbm{1}(d,u)$ is an indicator function representing the strategy of the news recommender system.
The indicator function $\mathbbm{1}(d,u)$ will output $1$ if the news $d$ is recommended to the user $u$ and will output $0$ if the news $d$ is not recommended to the user $u$.

\begin{figure*}
    \centering
    \resizebox{0.99\textwidth}{!}{
    \includegraphics{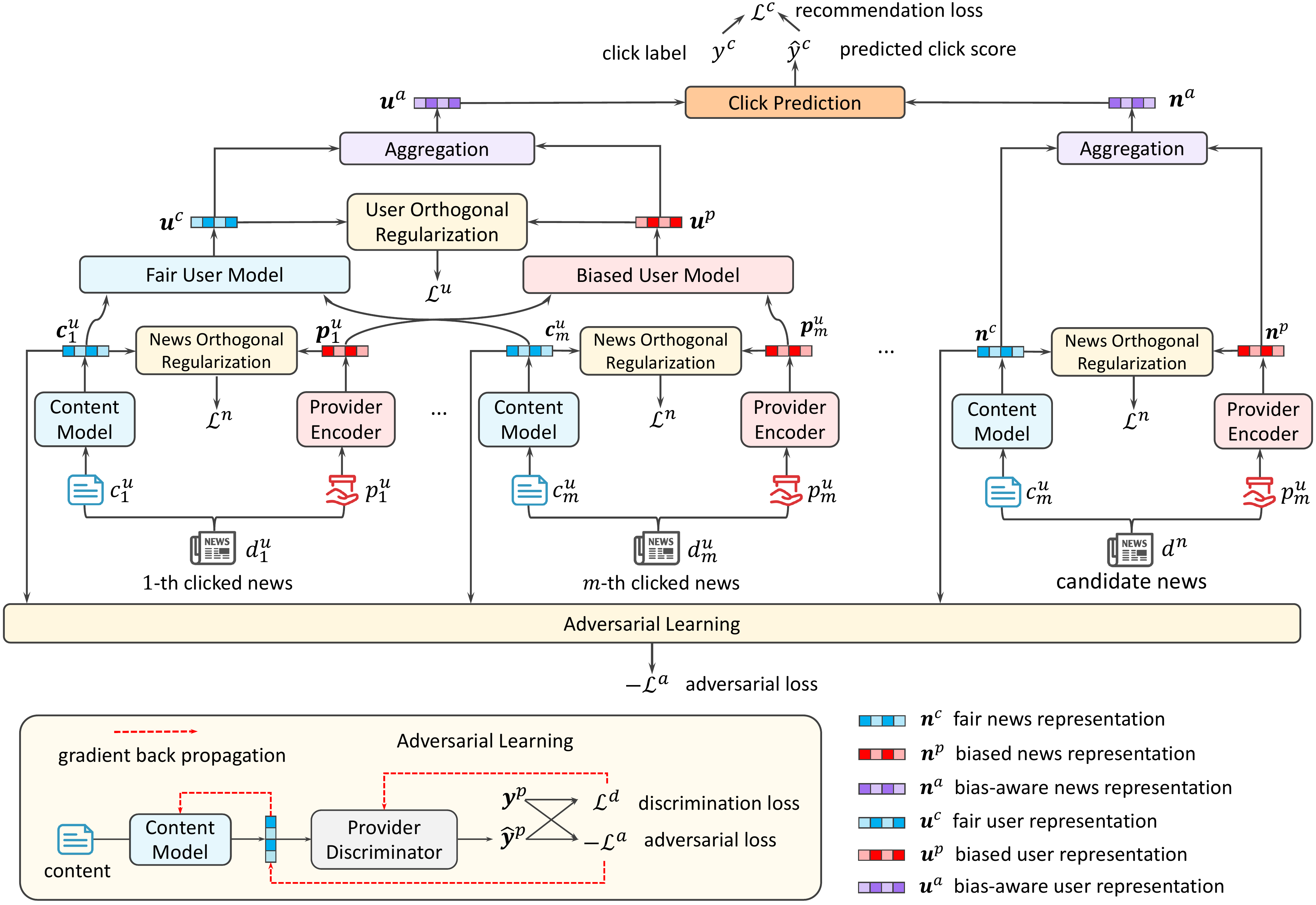}
    }
    \caption{The overall framework of \textit{ProFairRec}.}
    \label{fig.training}
\end{figure*}

\subsection{Overall Framework}
\label{sec.framework}

Next, we will introduce the overall framework of our provider fairness-aware news recommendation framework (\textit{ProFairRec}), which can learn fair news recommendation models from biased data.

As shown in Fig.~\ref{fig.training}, the core idea of \textit{ProFairRec} is to learn provider-fair news and user representations to achieve provider fairness.
Note that training data usually contains both provider-biased signals (e.g., provider popularity) and provider-independent signals (e.g., the matching of user interest and news content), it may be difficult to employ a single news or user representation to purely encode provider-independent signals without encoding provider-biased signals.
To tackle this challenge, motivated by \citet{wu2021fairness}, we employ provider-biased representations to inherit provider bias from biased data to further learn provider-fair representations.
Specifically, given a news $d$ with its content $c$ and provider ID $p$, \textit{ProFairRec} employs a content model $\Phi^n_c$ to learn provider-fair news representation $\textbf{c}$ from its content $c$, and a provider encoder $\Phi^n_p$ to learn provider-biased news representation $\textbf{p}$ from its provider ID p:
\begin{equation}
    \textbf{c} = \Phi_c^n(c), \quad \quad \textbf{p} = \Phi_p^n(p).
\end{equation}
The content model can be implemented by news modeling techniques proposed in existing methods~\cite{wu2019ijcai,qi2021pprec,wu2019neuralc}\footnote{We will introduce the architecture of the content model in Section~\ref{sec.model}.} and the provider encoder is implemented by the stack of a provider embedding layer and an MLP network.
Based on this decomposition, the provider-biased news representations have the potential to encode provider bias in training data and provider-fair news representations have the potential to purely encode bias-independent information.
Next, given a target user $u$, we apply the content model and the provider encoder to her clicked news to obtain the corresponding provider-fair news representations $[\textbf{c}^u_1,...,\textbf{c}^u_m]$ and provider-biased news representations $[\textbf{p}^u_1,...,\textbf{p}^u_m]$, where $\textbf{c}^u_i$ and $\textbf{p}^u_i$ denote provider-fair and -biased representations of the $i$-th clicked news $d^u_i$, respectively.
Given a candidate news $d^n$, we also apply the content model and the provider encoder to learn its provider-fair news representation $\textbf{n}^c$ and provider-biased news representation $\textbf{n}^p$.

Next, for user modeling, we also employ provider-biased representations to inherit provider-biased signals from training data to learn provider-fair representations to capture user interest. 
We employ a fair user model $\Phi^u_c$ to learn provider-fair user representation $\textbf{u}^c$ from provider-fair representations of user's clicked news:
\begin{equation}
    \textbf{u}^c = \Phi^u_c([\textbf{c}^u_1,...,\textbf{c}^u_i,...,\textbf{c}^u_m]),
\end{equation}
where the fair user model can be implemented by user modeling methods proposed in existing works~\cite{an2019neural,qi2021hierec,wu2021uag}\footnote{We will introduce the framework of the user model in Section~\ref{sec.model}.}.
We also employ a biased user model $\Phi^u_p$ to learn provider-biased user representation $\textbf{u}^p$ from provider-biased representations of user's clicked news:
\begin{equation}
\textbf{u}^p = \Phi^u_p([\textbf{p}^u_1,...,\textbf{p}^u_i,...,\textbf{p}^u_m]),
\end{equation}
where $\Phi^u_p$ is also implemented by existing user modeling methods.
Based on this decomposition, provider-biased user representations can encode biased signals in training data and provider-fair user representations could purely encode bias-independent information.

Since training data usually contains provider bias signals, provider-fair representations may inherit bias information if we only employ them for model training.
Thus, we use both provider-fair and -biased representations of users and news for training to fit the distribution of biased data.
Provider-biased representations are used to inherit bias from training data and provider-fair representations are used to capture bias-independent information.
In addition, to better prevent provider-fair news representations from encoding bias, we propose an adversarial learning task on news provider discrimination and apply it to provider-fair news representations.
We also propose an orthogonal regularization and apply it to the provider-fair and -biased representations of both user and news to better eliminate bias in provider-fair representations.
Thus, \textit{ProFairRec} has the ability to learn provider-fair representations from that purely encode bias-independent information from biased data.
Finally, we only use provider-fair news and user representations for interest matching to achieve provider fairness-aware news recommendation.


\subsection{Model Training}

Next, we will introduce the training framework of \textit{ProFairRec} in detail.
Following existing works~\cite{wu2019neuralc,an2019neural}, we also train news recommendation models on historical user click data.
As mentioned above, user click behaviors on news providers are usually biased and training data may encode bias on news providers.
News recommendation models trained on biased data may inherit provider bias and become unfair for minority providers.
To tackle this challenge, during model training of \textit{ProFairRec}, both provider-fair and provider-biased representations are used for click prediction to fit the distribution of biased training data.
The provider-biased representations are used to inherit provider bias from training data and provider-fair representations are used to encode bias-independent information.
Specifically, we first aggregate provider-fair and -biased representation of user and news to learn bias-aware user representation $\textbf{u}^a$ and bias-aware news representation $\textbf{n}^a$: $\textbf{u}^a = \textbf{u}^c + \textbf{u}^p, \textbf{n}^a = \textbf{n}^c + \textbf{n}^p.$
Next, we predict click score $\hat{y}^c$ of the target user $u$ on the candidate news $d^n$ based on the matching of their bias-aware representations: $\hat{y}^c=\textbf{u}^a\cdot \textbf{n}^a$.
Following existing works~\cite{wu2019ijcai,wu2019npa}, we apply the noise contrastive estimation (NCE) technique to formulate the recommendation loss $\mathcal{L}^c$.
We treat each news $d^+$ clicked by the user as a positive sample, and randomly select $F$ negative samples $[d^-_1,...,d^-_F]$ for each positive sample, where negative samples are non-clicked news in the same impression with the positive sample and $d^-_j$ denotes the $j$-th negative sample.
Next we calculate the click score $\hat{y}^c_+$ of the positive sample $d^+$ and click scores $[\hat{y}^c_1,...,\hat{y}^c_F]$ of the negative samples.
Finally, we obtain the recommendation loss $\mathcal{L}^c$ based on the NCE loss:
\begin{equation}
    \mathcal{L}^c  =\mathbb{E}_{x\sim \Omega}[ -\log(\frac{\exp(\hat{y}^c_+)}{\exp(\hat{y}_+^c)+\sum_{j=1}^F \exp(\hat{y}^c_j)} )],
\end{equation}
where $\Omega$ is the training dataset, $x$ is a training sample that is composed of a positive sample and $F$ negative samples.

Although in \textit{ProFairRec} we employ provider-biased representations to inherit provider bias in training data, provider-fair representations may also encode some bias information.
To protect provider-fair representations from encoding provider bias, we propose an adversarial learning task to enforce provider-fair news representations to be indistinguishable to different news providers.
We first employ a provider discriminator $\Psi(\cdot)$ to predict provider of a news $d$ from its provider-fair representation $\textbf{c}$, i.e., $\hat{\textbf{y}}^p = \Psi(\textbf{c})$, where $\hat{\textbf{y}}^p \in \mathbb{R}^{|\mathcal{P}|}$ denotes the predicted probability vector and its $j$-th element is the predicted probability of the news $d$ belonging to the $j$-th provider $P_j$.
We use an MLP network to implement the provider discriminator $\Psi(\cdot)$:
\begin{equation}
    \hat{\textbf{y}}^p = softmax (\textbf{W}_d\widetilde{\textbf{c}} + \textbf{b}_d), \quad \widetilde{\textbf{c}} = MLP(\textbf{c}),
\end{equation}
where $\textbf{W}_d$ and $\textbf{b}_d$ are trainable parameters, and $MLP(\cdot)$ is an MLP network.
Next, we formulate the provider discrimination loss $\mathcal{L}^d$ and minimize $\mathcal{L}^d$ to learn the optimal provider discriminator $\Psi^*$:
\begin{equation}
  \Psi^* = \arg\min_{\Psi} \mathcal{L}^d, \quad  \mathcal{L}^d = \mathbb{E}_{d\in \mathcal{D}} [- \textbf{y}^p\cdot\log\hat{\textbf{y}}^p],
\end{equation}
where $\textbf{y}^p$ is the label of the provider of news $d$.
After obtaining the optimal provider discriminator $\Psi^*$, we recalculate $\hat{\textbf{y}}^p$ based on $\Psi^*$ and further calculate the adversarial loss $\mathcal{L}^a$:
\begin{equation}
\label{eq.adv}
    \mathcal{L}^a = \mathbb{E}_{d\in \mathcal{D}} [- \textbf{y}^p\cdot\log\hat{\textbf{y}}^p | \Psi^*].
\end{equation}
Based on this equation, we can calculate and employ the negative adversarial gradients on the recommendation model as a penalty to reduce provider bias encoded in provider-fair news representations.

Besides, to achieve a better decomposition of provider bias and bias-independent information, we also apply an orthogonal regularization to provider-fair and -biased representations of both news and users.
For each news or user, orthogonal regularization constrains its provider-fair representation to be orthogonal with the corresponding provider-biased representation.
We first calculate cosine similarities between provider-fair and -biased representations, and then calculate the orthogonal regularization:
\begin{equation}
    \mathcal{L}^u = \mathbb{E}_{u\in \mathcal{U}} [|s^u|], \quad s^u = \frac{\textbf{u}^c\cdot \textbf{u}^p}{||\textbf{u}^c||_2 ||\textbf{u}^p||_2},
\end{equation}
\begin{equation}
    \mathcal{L}^n = \mathbb{E}_{d\in \mathcal{D}} [|s^n|], \quad s^n = \frac{\textbf{n}^c\cdot \textbf{n}^p}{||\textbf{n}^c||_2 ||\textbf{n}^p||_2},
\end{equation}
where $s^u$ is the cosine similarity between provider-fair and -biased representation of the user $u$, $s^n$ is the cosine similarity between provider-fair and -biased representation of the news $d$, $\mathcal{L}^u$ and $\mathcal{L}^n$ denote the user and news orthogonal regularization, respectively.

Finally, we combine different loss functions into an overall optimization objective $\mathcal{L}$ for model training:
\begin{equation}
    \mathcal{L} = \lambda_c \mathcal{L}^c + \lambda_u \mathcal{L}^u + \lambda_n \mathcal{L}^n - \lambda_a \mathcal{L}^a,
\end{equation}
where $\lambda_c$, $\lambda_u$, $\lambda_n$, $\lambda_a$ are hyper-parameters controlling the weights of loss $\mathcal{L}^c$, $\mathcal{L}^u$, $\mathcal{L}^n$ and $\mathcal{L}^a$, respectively.

\subsection{Fair News Recommendation}

Next, we will introduce the fair recommendation framework of \textit{ProFairRec}.
After training the recommendation model on user data, provider bias is enforced to be encoded in provider-biased representations and be reduced from provider-fair representations.
Since provider-fair representations only encode bias-independent information, we can match provider-fair news and user representations to achieve fair news recommendation.
Thus, given a target user $u$ and a candidate news $d^n$, based on the content model and fair user model, we first learn provider-fair news representation $\textbf{n}^c$ for the candidate news $d^n$ and provider-fair user representation $\textbf{u}^c$ for the target user $u$.
We further match them to model user interest in candidate news without being affected by provider bias: $\hat{r}^c = \sigma(\textbf{u}^c\cdot \textbf{n}^c),$ where $\hat{r}^c$ is the predicted unbiased click probability and $\sigma(\cdot)$ is the sigmoid activation function.
Finally, given a target user $u$ and a list of candidate news $[d^n_1,...,d^n_i,...,d^n_N]$, we can predict unbiased click probability for each candidate news and further rank these candidate news for recommendation.

\subsection{Recommendation Model}
\label{sec.model}

As mentioned in Section~\ref{sec.framework}, the content model and the user model of \textit{ProFairRec} can be implemented by news modeling and user modeling techniques proposed in existing methods.
Thus, \textit{ProFairRec} is model-agnostic and can be combined with existing news recommendation models to improve their provider fairness.
Next, we will briefly introduce the architecture of mainstream news recommendation models.
These models usually contain a content model to learn news representations from textual news content and a user model to learn user representations from users' reading history.

The content model contains three major components, i.e., a word embedding layer, a textual context modeling network, and a text pooling network.
Given words in news content, they are first converted into a word embedding via the word embedding layer.
Next, a textual context modeling network is applied to the word embedding sequence to capture relatedness among words and learn contextual representations for them.
This text context modeling network can be implemented by various NLP techniques, such as CNN network~\cite{lecun1998gradient,an2019neural}, LSTM network~\cite{hochreiter1997long,wu2020mind}, and transformer network~\cite{vaswani2017attention,ge2020graph}.
Finally, contextual word representations are aggregated by the text pooling network to obtain a news representation $\textbf{d}$.
The text pooling network can be implemented by various pooling operations, such as average pooling~\cite{wang2018dkn} and attention network~\cite{wu2019ijcai}.

The user model usually contains two major components, i.e., a behavior context modeling network and a behavior pooling network.
User's clicked news are first converted to news embeddings via the content model.
Then we apply a behavior context modeling network to understand user interest from user behavior contexts.
It takes embeddings of user's clicked news as input and generates contextual news representations.
This layer is usually implemented by some effective sequence modeling techniques, such as GRU network~\cite{okura2017embedding,an2019neural} and multi-head self-attention network~\cite{wu2019neuralc}. 
Next, we apply the behavior pooling network to build user interest representation $\textbf{u}$.
It is usually implemented by attention network~\cite{wu2019ijcai,wu2019neuralc}.

\section{experiment}
\subsection{Datasets and Experimental Settings}

We conduct extensive experiments on a public news recommendation dataset~\cite{wu2020mind} (\textit{MIND})\footnote{https://msnews.github.io/index.html.} to evaluate model performance and fairness.
\textit{MIND} is constructed by click data of 1 million uss on the Microsoft News\footnote{https://www.msn.com/en-us/news/us} from October 12 to November 22, 2019 (6 weeks).
User data in the first four weeks are used to construct click history, user data in the last week are used to construct the test set and other user data are used to construct the training and validation set.
Besides, since \textit{MIND} does not contains provider information, we extract them from the website via a web crawler.
More details of the \textit{MIND} dataset are in Table~\ref{table.stat}.

\begin{table}[h]
\centering
\caption{Details of the \textit{MIND} dataset. }
\label{table.stat}
\resizebox{0.4\textwidth}{!}{
\begin{tabular}{cccc}
\Xhline{1pt}
\# News       & 130,379      & \# Providers      & 1,705       \\
\# Users      & 1,000,000     & \# Impressions    & 4,979,946   \\
\# Clicks     & 7,583,733   & \# Non-clicks     & 183,124,199 \\
\multicolumn{3}{c}{Avg. \# words in news titles}  & 11.78        \\ \Xhline{1pt}
\end{tabular}
}
\end{table}

Next, we will introduce hyper-parameter settings of \textit{ProFairRec}.
Following \citet{wu2019neuralc,wu2019npa}, we utilize the first $30$ words in news titles for news content modeling and users' recent $50$ clicked news for user modeling.
We select a famous news recommendation method, i.e., \textit{NRMS}~\cite{wu2019neuralc}, to implement the content and user model of \textit{ProFairRec}.
The content model is based on the stack of a word embedding lay, a multi-head self-attention (MHSA) network, and an attention pooling network. 
Specifically, the word embedding layer is initialized by pre-trained 300-dimensional Glove word embeddings~\cite{pennington2014glove} and fine-tuned in experiments.
The MHSA network is set to contain $20$ attention heads, and the output dimension of each head is set to $20$.
The attention network is implemented by a two-layer MLP network.
The user model is based on the stack of an MHSA network and an attention pooling network.
Similarly, the MHSA generates $400$-dimensional output vectors, and the attention pooling network is a two-layer MLP network.
Besides, for the provider encoder, the 400-dimensional provider embeddings are randomly initialized and fine-tuned in experiments.
We randomly select $4$ negative samples for each positive sample to formulate the recommendation loss $\mathcal{L}^c$ and set its weight $\lambda_c$ to 1.
Both the weights $\lambda_u$ and $\lambda_n$ of news and user orthogonal regularization $\mathcal{L}^u$ and $\mathcal{L}^n$ are set to 1.
In addition, since there are many providers in \textit{MIND} which brings difficulties to the convergence of the adversarial learning, we retain 50 providers with the most clicks and merge other providers into a single class category for discrimination.
Besides, following \citet{goodfellow2020generative}, we iteratively train discrimination and adversarial task for a single step rather than finding the optimal discriminator in practice.
The weight $\lambda_a$ of the adversarial loss $\mathcal{L}^a$ is set to 0.004.
We utilize the Adam algorithm~\cite{kingma2014adam} with $0.0001$ learning rate for model optimization.
We select hyper-parameters based on the validation dataset.
Codes are released at \url{https://github.com/taoqi98/ProFairRec}.

\begin{table}[]
\caption{Comparisons of \textit{ProFairRec} with baseline methods on recommendation performance.}
\label{table.performance}
\centering
\resizebox{0.47\textwidth}{!}{
\begin{tabular}{cccc}
\Xhline{1.2pt}
           & AUC            & MRR            & nDCG@10        \\ \hline
GRU~\cite{okura2017embedding}        & 66.50$\pm$0.04 & 32.06$\pm$0.04 & 40.45$\pm$0.02 \\
DKN~\cite{wang2018dkn}        & 66.70$\pm$0.15 & 32.41$\pm$0.11 & 40.88$\pm$0.12 \\
HiFiArk~\cite{liu2019hi}    & 67.49$\pm$0.19 & 33.04$\pm$0.15 & 41.57$\pm$0.15 \\
NAML~\cite{wu2019ijcai}       & 67.22$\pm$0.20 & 33.01$\pm$0.10 & 41.54$\pm$0.12 \\
NPA~\cite{wu2019npa}        & 67.13$\pm$0.07 & 32.90$\pm$0.07 & 41.45$\pm$0.07 \\
KRED~\cite{liu2020kred}       & 67.55$\pm$0.11 & 33.27$\pm$0.05 & 41.83$\pm$0.05 \\
GNewsRec~\cite{hu2020graph}   & 68.41$\pm$0.10 & 33.59$\pm$0.10 & 42.24$\pm$0.11 \\
LSTUR~\cite{an2019neural}      & 68.35$\pm$0.10 & 33.48$\pm$0.10 & 42.16$\pm$0.09 \\
NRMS~\cite{wu2019neuralc}       & 68.08$\pm$0.13 & 33.43$\pm$0.09 & 42.07$\pm$0.10 \\ \hline
OFAiR~\cite{sonboli2020opportunistic}      & 67.46$\pm$0.17 & 33.08$\pm$0.14 & 41.62$\pm$0.17 \\
TFORM~\cite{wu2021tfrom}      & 67.53$\pm$0.16 & 33.12$\pm$0.13 & 41.63$\pm$0.16 \\ \hline
ProFairRec & 67.64$\pm$0.10 & 33.08$\pm$0.05 & 41.67$\pm$0.07 \\

\Xhline{1.2pt}
\end{tabular}
}
\end{table}

\begin{figure*}
    \centering
    \resizebox{0.99\textwidth}{!}{
    \includegraphics{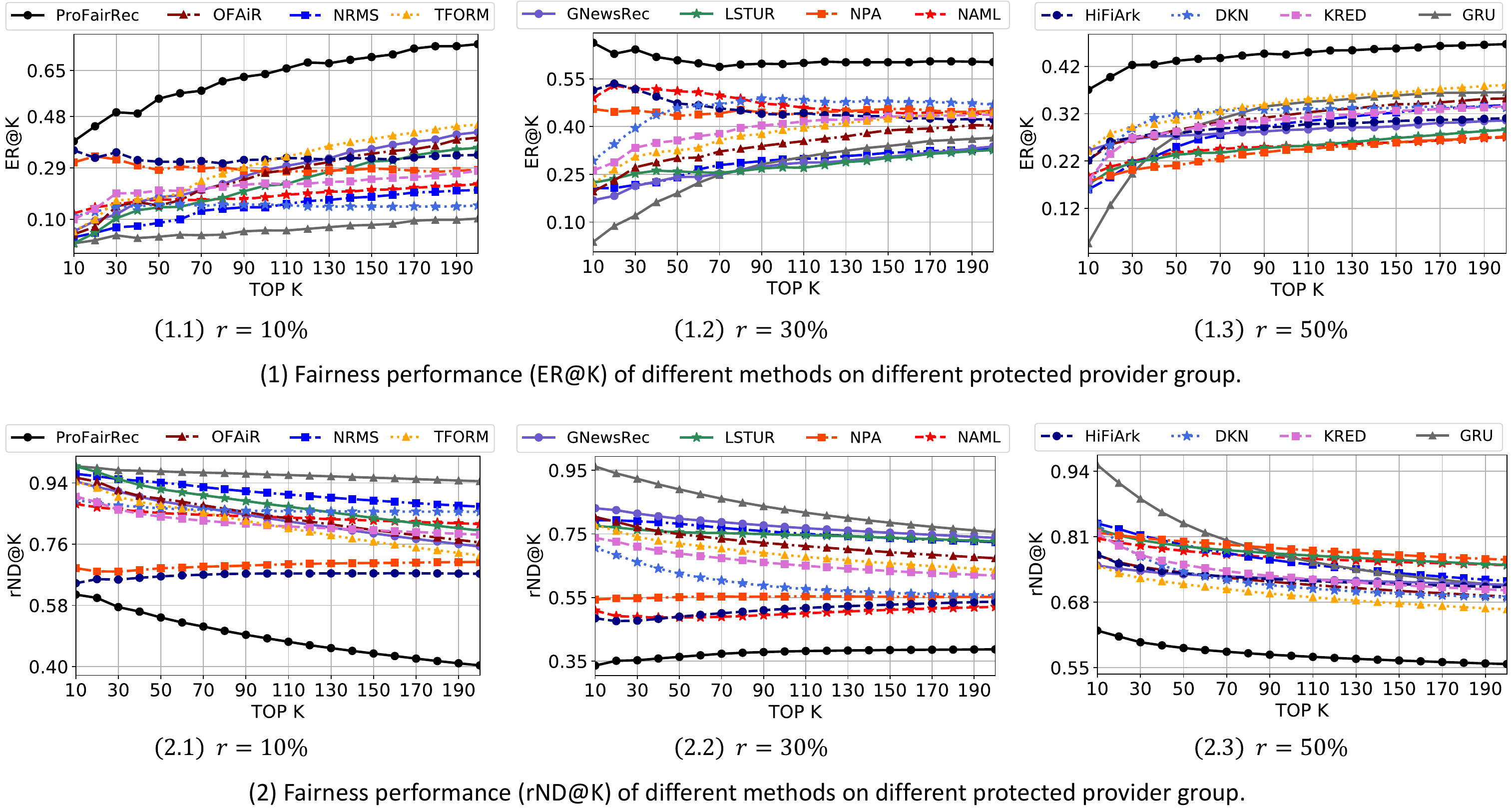}
    }
    \caption{Provider fairness of \textit{ProFairRec} and baseline methods based on different partition of protected provider group. Fairness is better if \textit{ER@K} is closer to \textit{1} and \textit{rND@K} is closer to \textit{0}.}
    \label{fig.mainfair}
\end{figure*}

Following existing works~\cite{an2019neural,wu2020mind,ge2020graph}, we use \textit{AUC}, \textit{nDCG@10} and \textit{MRR} for recommendation evaluation.
Besides, following existing works~\cite{burke2018balanced,yang2017measuring}, group-level fairness metrics are used for fairness evaluation and provider popularity is used for group partition.
We evaluate the provider fairness based on the difference of exposure number of providers in a protected group and an unprotected group.
Providers are first ranked based on their average click number, where unpopular providers ranked in the bottom $r\%$ are partitioned into the protected group $\mathcal{P}^+$ while others are partitioned into the unprotected group $\mathcal{P}^-$.
Fairness is evaluated under different $r$: $10\%$, $30\%$ and $50\%$.
Following existing works~\cite{burke2018balanced,yang2017measuring}, we employs two metrics to measure the exposure difference of providers in different groups based on the top $K$ recommended news, i.e., relative exposure ratio \textit{ER@K} and normalized discounted difference \textit{rND@K}:
\begin{equation}
\label{eq.}
    ER@K = \frac{\mathbb{E}_{u\in\mathcal{U}} [ |\mathcal{R}^u_K\cap\mathcal{D}^+|/|\mathcal{D}^+| ] }{\mathbb{E}_{u\in\mathcal{U}} [|\mathcal{R}^u_K\cap \mathcal{D}^-|/|\mathcal{D}^-|] },
\end{equation}
\begin{equation}
\label{eq.rND}
    rND@K = \frac{1}{Z} \mathbb{E}_{u\in \mathcal{U}}[\sum_{n=10,20,...}^K \frac{1}{log_2n}|\frac{|\mathcal{R}^u_n\cap\mathcal{D}^+|}{n}   -\frac{|\mathcal{D}^+|}{|\mathcal{D}|}| ],
\end{equation}
\begin{equation}
    \mathcal{D}^+ = \bigcup\limits_{p\in \mathcal{P}^+} \mathcal{D}_p, \quad \mathcal{D}^- = \bigcup\limits_{p\in \mathcal{P}^-} \mathcal{D}_p,
\end{equation}
where $\mathcal{R}^u_K$ is the top $K$ news recommended to the user $u$ and $Z$ is the normalization factor.
Eq.~\ref{eq.} and Eq.~\ref{eq.rND} show that the model is fairer if $ER@K$ is closer to $1$ or $rND@K$ is closer to $0$.
Besides, we rank all news in the dataset to compare the fairness of different methods, since original news impression may be biased.

\subsection{Performance Evaluation}

In this section, we compare \textit{ProFairRec} with baseline methods on both recommendation performance and provider fairness.
The first group of baseline methods includes several mainstream news recommendation methods:
(1) \textit{GRU}~\cite{okura2017embedding}: using a text auto-encoder to learn news representations and a GRU network to learn user representations.
(2) \textit{DKN}~\cite{wang2018dkn}: proposing a knowledge-aware CNN network to learn news representations from news texts and entities.
(3) \textit{HiFi-Ark}~\cite{liu2019hi}: proposing to learn multiple achieved user representations to model diverse user interest.
(4) \textit{NAML}~\cite{wu2019ijcai}: learning news representations from news titles, topics, subtopics, and bodies via an attentive CNN network.
(5) \textit{NPA}~\cite{wu2019npa}: proposing personalized attention networks to learn news representations from news titles and user representations by aggregating user's clicked news.
(6) \textit{KRED}~\cite{liu2020kred}: proposing a knowledge-aware graph attention network to learn news representations from news texts, entities, and neighbors of entities on knowledge graphs.
(7) \textit{GNewsRec}~\cite{hu2020graph}: modeling long-term user interest from news-user graph via a GNN network and short-term user interest from users' recent clicked news via a GRU network.
(8) \textit{LSTUR}~\cite{an2019neural}: learning user representations by combining short-term user interest inferred from reading history via a GRU network and long-term user interest modeled by user ID embeddings.
(9) \textit{NRMS}~\cite{wu2019neuralc}: proposing to use multi-head self-attention networks to learn news and user representations.

\begin{table*}[]
\caption{Provider fairness of different news recommendation methods and their combination with \textit{ProFairRec}. $r=50\%$.}
\label{table.general.fair}
\resizebox{0.99\textwidth}{!}{
\begin{tabular}{ccccccc}
\Xhline{1.2pt}
                  & ER@10                & ER@30                & ER@50                & rND@10                & rND@30                & rND@50                \\ \hline
NAML      & 0.1646$\pm$0.0080 & 0.2099$\pm$0.0107 & 0.2236$\pm$0.0102& 0.8316$\pm$0.0082 & 0.8130$\pm$0.0077 & 0.8022$\pm$0.0077\\
+ProFairRec & \textbf{0.2644}$\pm$0.0329 & \textbf{0.2721}$\pm$0.0229 & \textbf{0.2822}$\pm$0.0255& \textbf{0.7303}$\pm$0.0332 & \textbf{0.7271}$\pm$0.0261 & \textbf{0.7232}$\pm$0.0230\\ \hline
KRED      & 0.1518$\pm$0.0399 & 0.2271$\pm$0.0349 & 0.2572$\pm$0.0321& 0.8447$\pm$0.0407 & 0.8138$\pm$0.0379 & 0.7940$\pm$0.0356\\
+ProFairRec  & \textbf{0.2223}$\pm$0.0777 & \textbf{0.2627}$\pm$0.0649 & \textbf{0.2918}$\pm$0.0596& \textbf{0.7730}$\pm$0.0788 & \textbf{0.7565}$\pm$0.0734 & \textbf{0.7425}$\pm$0.0699\\ \hline
NPA       & 0.1503$\pm$0.0358 & 0.1894$\pm$0.0203 & 0.2086$\pm$0.0220& 0.8462$\pm$0.0364 & 0.8301$\pm$0.0298 & 0.8189$\pm$0.0276\\
+ProFairRec   & \textbf{0.2444}$\pm$0.0382 & \textbf{0.2499}$\pm$0.0361 & \textbf{0.2584}$\pm$0.0332& \textbf{0.7506}$\pm$0.0388 & \textbf{0.7483}$\pm$0.0373 & \textbf{0.7452}$\pm$0.0358\\ \hline
NRMS      & 0.1237$\pm$0.0243 & 0.1991$\pm$0.0213 & 0.2378$\pm$0.0195& 0.8734$\pm$0.0248 & 0.8423$\pm$0.0225 & 0.8202$\pm$0.0212\\
+ProFairRec  & \textbf{0.3644}$\pm$0.0847 & \textbf{0.3941}$\pm$0.0689 & \textbf{0.4156}$\pm$0.0658& \textbf{0.6295}$\pm$0.0854 & \textbf{0.6174}$\pm$0.0779 & \textbf{0.6071}$\pm$0.0738\\ \hline
LSTUR     & 0.1583$\pm$0.0206 & 0.2086$\pm$0.0292 & 0.2254$\pm$0.0326& 0.8381$\pm$0.0210 & 0.8174$\pm$0.0232 & 0.8050$\pm$0.0257\\
+ProFairRec& \textbf{0.3765}$\pm$0.1067 & \textbf{0.4491}$\pm$0.0791 & \textbf{0.4865}$\pm$0.0663& \textbf{0.6174}$\pm$0.1070 & \textbf{0.5879}$\pm$0.0952 & \textbf{0.5668}$\pm$0.0869\\ \hline
\Xhline{1.2pt}

\end{tabular}
}
\end{table*}

\begin{table}[]
\caption{Recommendation performance of baseline news recommendation methods and their combinations with our \textit{ProFairRec} approach.}
\label{table.general.performance}
\centering
\resizebox{0.45\textwidth}{!}{
\begin{tabular}{cccc}
\Xhline{1.2pt}
           & AUC            & MRR            & nDCG@10        \\ \hline
NAML      & 67.22$\pm$0.20 & 33.01$\pm$0.10 & 41.54$\pm$0.12 \\
+ProFairRec  & 67.13$\pm$0.08 & 32.86$\pm$0.09 & 41.38$\pm$0.09 \\ \hline
KRED      & 67.55$\pm$0.11 & 33.27$\pm$0.05 & 41.83$\pm$0.05 \\
+ProFairRec  & 67.51$\pm$0.19 & 33.11$\pm$0.16 & 41.71$\pm$0.16 \\ \hline
NPA       & 67.13$\pm$0.07 & 32.90$\pm$0.07 & 41.45$\pm$0.07 \\
+ProFairRec   & 67.13$\pm$0.03 & 32.86$\pm$0.05 & 41.39$\pm$0.05 \\ \hline
NRMS      & 68.08$\pm$0.13 & 33.43$\pm$0.09 & 42.07$\pm$0.10 \\
+ProFairRec  & 67.64$\pm$0.10 & 33.08$\pm$0.05 & 41.67$\pm$0.07 \\ \hline
LSTUR     & 68.35$\pm$0.10 & 33.48$\pm$0.10 & 42.16$\pm$0.09 \\
+ProFairRec & 67.46$\pm$0.11 & 32.83$\pm$0.13 & 41.35$\pm$0.13 \\ \hline
\Xhline{1.2pt}
\end{tabular}
}
\end{table}

The second group of baseline methods includes several provider fairness-aware recommendation methods.
(1) \textit{OFAiR}~\cite{sonboli2020opportunistic}: re-ranking items based on personalized relevance and provider unfairness of recommendation results.
(2) \textit{TFORM}~\cite{wu2021tfrom}: dynamically re-balancing the exposure opportunities of providers based on their historical exposures.
Note that these methods apply re-ranking techniques to improve the provider fairness in item recommendation and cannot be directly applied to the news recommendation task, thus we apply the core re-ranking techniques in them to re-rank recommendation results of \textit{NRMS} as baseline provider fairness-aware news recommendation methods for comparisons.
Besides, since the trade-off between recommendation performance and fairness of these two baseline methods and \textit{ProFairRec} can be adjusted via hyper-parameters, we compare the provider fairness of them under similar recommendation performance for fair comparisons.
We repeat each experiment five times and report average results.

Table~\ref{table.performance} shows the recommendation performance of different methods and Fig.~\ref{fig.mainfair} shows the provider fairness of different methods.
From these results, we have several findings.
First, baseline news recommendation methods tend to recommend news of popular providers.
This is because user click data on news providers are usually biased.
Recommendation models trained on biased data may learn some shortcuts on news providers for recommendation and may be unfair for some minority news providers.
Second, our \textit{ProFairRec} method outperforms baseline news recommendation methods on provider fairness.
This is because we propose to decompose provider bias and bias-independent information of user data into provider-biased and -fair representations and only employ provider-fair representations for news ranking.
Besides, we also propose an adversarial learning task on provider discrimination and orthogonal regularization to achieve better bias decomposition.
Third, our \textit{ProFairRec} method effectively outperforms baseline fair recommendation methods on provider fairness when their recommendation performance are similar.
This is because existing provider fairness-aware recommendation methods are usually based on re-ranking techniques.
These methods independently optimize model performance and fairness and may only achieve a sub-optimal trade-off between performance and fairness.
Different from these methods, our \textit{ProFairRec} method jointly optimizes recommendation performance and provider fairness based on adversarial decomposition technique, which can more effectively improve provider fairness of news recommendation models.
Forth, compared with baseline news recommendation methods, our \textit{ProFairRec} method can achieve comparable or even better recommendation performance.
These results validate that our \textit{ProFairRec} method can effectively improve recommendation fairness and meanwhile accurately recommend users' interested news as existing news recommendation methods.

\subsection{Generalization of \textit{ProFairRec}}
\label{sec.general}

As mentioned above, \textit{ProFairRec} is a general framework and we can combine \textit{ProFairRec} with models of existing news recommendation methods to improve their fairness.
Thus, to verify the generalization of \textit{ProFairRec}, we apply \textit{ProFairRec} to several news recommendation models that achieve good recommendation performance in Table~\ref{table.performance}, i.e., \textit{NAML}, \textit{KRED}, \textit{NPA}, \textit{NRMS} and \textit{LSTUR} and evaluate their fairness and performance.
Specifically, we first employ the adversarial training framework of \textit{ProFairRec} to learn fair user and news models for these methods.
Then we employ the fair recommendation framework of \textit{ProFairRec} for recommendation based on the fair user and news models in these methods.
Results on provider fairness are summarized in Table~\ref{table.general.fair} and results on recommendation performance are summarized  in Table~\ref{table.general.performance}.
Due to space limitations, we only show results under the protected provider group with $r=50\%$.
From these results, we find that our \textit{ProFairRec} approach can effectively and consistently improve the provider fairness of existing news recommendation models with minor performance decline.
This is because our \textit{ProFairRec} approach employs provider-biased representations to inherit provider-bias signals from the biased training data and thereby can learn provider-fair representations via adversarial learning for both user and news to achieve provider fairness in news recommendation.
In addition, our \textit{ProFairRec} approach is a model-agnostics framework, which can be combined with existing news recommendation methods to improve their fairness.
These results further validate the effectiveness and generalization of \textit{ProFairRec} on improving provider fairness.

\subsection{Ablation Study}



\begin{figure}
    \centering
    \resizebox{0.5\textwidth}{!}{
    \includegraphics{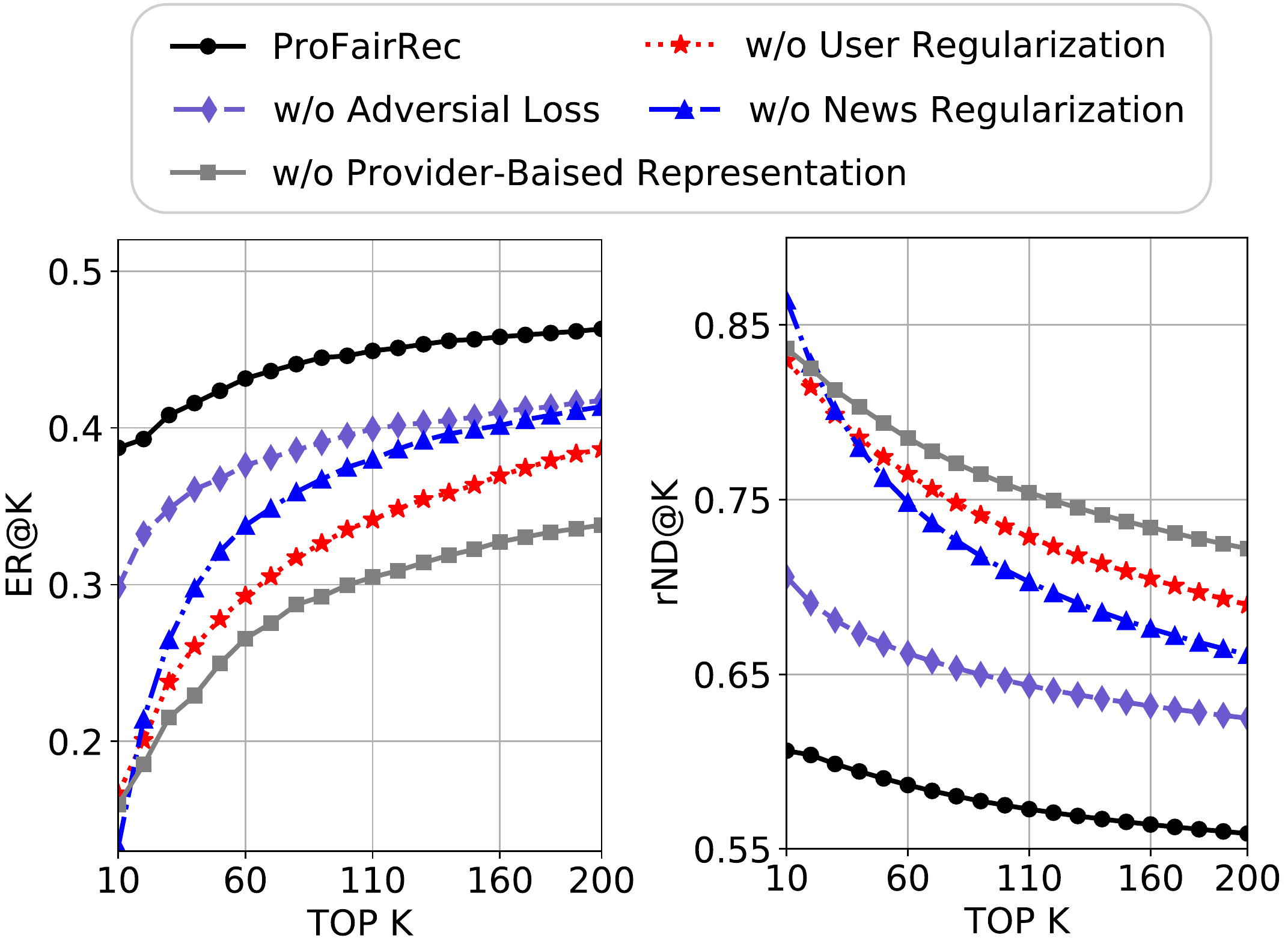}
    }
      \caption{Ablation analysis of our \textit{ProFairRec} approach.}
    \label{fig.ablation}
\end{figure}

Next, we conduct an ablation study to evaluate the effectiveness of several important modules in \textit{ProFairRec}, i.e., provider-biased representations, adversarial learning, and orthogonal regularization, on the provider fairness by removing them individually.
Results are summarized in Fig.~\ref{fig.ablation}, from which we have three findings.
First, after removing the provider-biased representations, provider fairness of \textit{ProFairRec} significantly declines.
This is because training data usually includes both provider bias and bias-independent signals, and it is difficult for a single news or user representation to purely encode bias-independent information without encoding provider bias.
To tackle this challenge, \textit{ProFairRec} decomposes provider bias and bias-independent information of biased data into provider-biased and -fair representations respectively, and only employ provider-fair representations for recommendation to improve provider fairness.
Second, adversarial learning can also improve the provider fairness of \textit{ProFairRec}.
This is because provider-fair representations may also inherit some bias information from biased training data.
The adversarial learning enforces provider-fair news representations to be indistinguishable with respect to different news providers, which can further prevent provider-fair news representations from encoding provider bias.
Third, removing the orthogonal regularization also hurts the provider fairness of \textit{ProFairRec}.
This may be because the orthogonal regularization constrains the orthogonality of provider-biased and -fair representations, which can better decompose provider bias and bias-independent information.

\subsection{Influence of Hyper-Parameters}

\begin{figure}
    \centering
    \resizebox{0.5\textwidth}{!}{
    \includegraphics{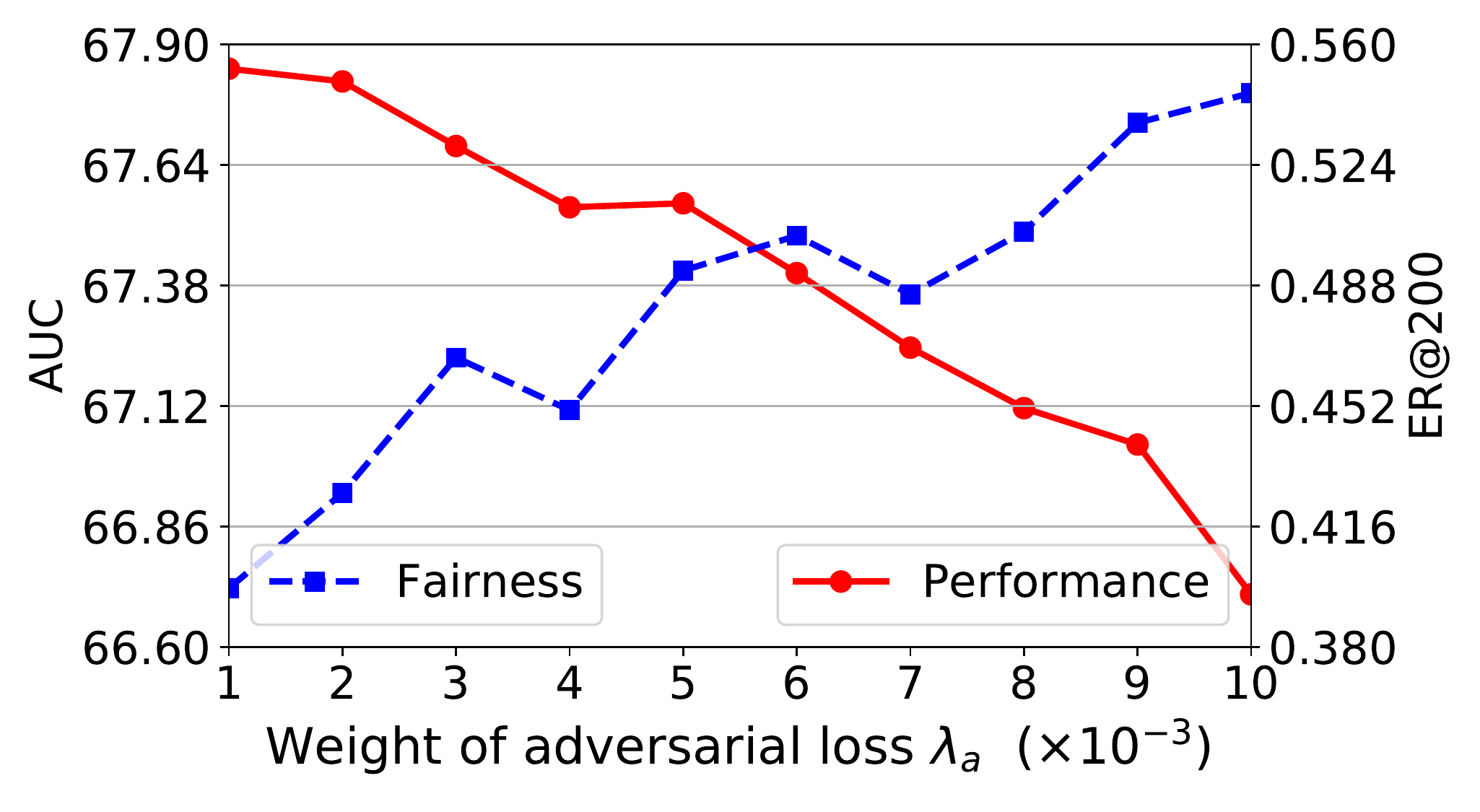}
    }
    \caption{Influence of adversarial learning loss weight $\lambda_a$ on recommendation performance and fairness.}
    \label{fig.hyp}
\end{figure}

In this section, we will explore how the adversarial learning in \textit{ProFairRec} affects the recommendation performance and provider fairness.
We analyze its influence by evaluating \textit{ProFairRec} under different weights of the adversarial learning loss $\lambda_a$.
Results are summarized in Fig.~\ref{fig.hyp}, from which we have two major findings.
First, with the increase of $\lambda_a$, the recommendation fairness of \textit{ProFairRec} consistently increases.
This is intuitive since larger $\lambda_a$ makes it more important for the model to reduce news provider information in provider-fair news representations, which can effectively improve the provider fairness.
Second, large $\lambda_a$ hurts the recommendation performance of \textit{ProFairRec}.
This is because large $\lambda_a$ makes the recommendation model overly emphasize the adversarial learning task while the recommendation task cannot get enough respect.
Thus, by adjusting $\lambda_a$, we can achieve a trade-off between recommendation performance and provider fairness.

\section{CONCLUSION}

In this paper, we propose a provider fairness-aware news recommendation framework (\textit{ProFairRec}), which can learn fair news recommendation models from biased user data.
The core of \textit{ProFairRec} is to learn provider-fair representations for both news and users to encode bias-independent information for recommendation.
Specifically, we propose to learn provider-fair and -biased representations for news from their content and provider IDs respectively.
These representations are further aggregated to build provider-fair and -biased user representations based on user reading history.
Both provider-fair and -biased representations are used for model training while only provide-fair representations are used for recommendation to improve provider fairness.
In addition, we apply adversarial learning to provider-fair news representations to prevent them from encoding bias.
We also apply orthogonal regularization to provider-fair and -biased representations to better reduce bias in provider-fair representations.
Besides, \textit{ProFairRec} is a model-agnostics framework and can be applied to existing news recommendation models to improve their fairness.
Experiments on a public dataset demonstrate that \textit{ProFairRec} can effectively improve the fairness of mainstream news recommendation methods with minor performance declines.
In our future work, we plan to apply \textit{ProFairRec} to improve provider fairness in more recommendation scenarios (e.g. item recommendation).


\section*{Acknowledgments}
This work was supported by the National Natural Science Foundation of China under Grant numbers 2021ZD0113902, U1936208, U1936126 and Tsinghua-Toyota Joint Research Funds 20213930033.

\bibliographystyle{ACM-Reference-Format}
\bibliography{mybib}


\end{document}